\documentclass[english,reprint, longbibliography, superscriptaddress, breaklinks=true, showkeys, showpacs=false, nofootinbib]{revtex4}
\usepackage[T1]{fontenc}
\usepackage[latin9]{inputenc}
\usepackage{color}
\usepackage{babel}
\usepackage{amsmath}
\usepackage{amssymb}
\usepackage{subfigure}
\usepackage{graphicx}
\usepackage{physics}
\usepackage[unicode=true,pdfusetitle,
 bookmarks=true,bookmarksnumbered=false,bookmarksopen=false,breaklinks=true,pdfborder={0 0 0},backref=false,colorlinks=true]
 {hyperref} 
\makeatletter
\@ifundefined{textcolor}{}
{%
 \definecolor{BLACK}{gray}{0}
 \definecolor{WHITE}{gray}{1}
 \definecolor{RED}{rgb}{1,0,0}
 \definecolor{GREEN}{rgb}{0,1,0}
 \definecolor{BLUE}{rgb}{0,0,1}
 \definecolor{CYAN}{cmyk}{1,0,0,0}
 \definecolor{MAGENTA}{cmyk}{0,1,0,0}
 \definecolor{YELLOW}{cmyk}{0,0,1,0}
}
\pdfoutput=1
\hypersetup{colorlinks=true,citecolor=blue,linkcolor=cyan,urlcolor=blue,filecolor= green, breaklinks=true}
\usepackage{url}
\usepackage{breakurl}
\makeatother

\begin{document}

\title{Quantum coherence versus interferometric visibility \\ in a biased Mach-Zehnder interferometer}

\author{Diego S. S. Chrysosthemos}
\email{starkediego@gmail.com}
\address{Departamento de F\'isica, Centro de Ci\^encias Naturais e Exatas, Universidade Federal de Santa Maria, Avenida Roraima 1000, Santa Maria, Rio Grande do Sul, 97105-900, Brazil}

\author{Marcos L. W. Basso}
\email{marcoslwbasso@hotmail.com}
\address{Centro de Ci\^encias Naturais e Humanas, Universidade Federal do ABC, Avenida dos Estados 5001, Santo Andr\'e, S\~ao Paulo, 09210-580, Brazil}

\author{Jonas Maziero}
\email{jonas.maziero@ufsm.br (corresponding author)}
\address{Departamento de F\'isica, Centro de Ci\^encias Naturais e Exatas, Universidade Federal de Santa Maria, Avenida Roraima 1000, Santa Maria, Rio Grande do Sul, 97105-900, Brazil}

\selectlanguage{english}%

\begin{abstract}
The double-slit interferometer and the Mach-Zehnder interferometer (MZI) with balanced beam splitters are prototypical setups for investigating the quantum wave-particle duality. These setups induced a quantitative association of interferometric visibility (IVI) with the wave aspect of a single quantum system (WAQ). Recently, it was realized that quantum coherence (QC) can be better suited than IVI for quantifying the WAQ in complementarity relations. In this article, we investigate a MZI with biased beam splitters both in the input and the output, and we show that in some cases the IVI is not adequate to quantify the WAQ since it does not reflect the behavior of the quantum coherence, even for a bi-dimensional closed quantum system.  Using IBM quantum computers, we experimentally verify our theoretical findings by doing a full quantum simulation of the optical MZI with biased beam splitters.
\end{abstract}


\date{\today}

\maketitle

\section{Introduction}
The wave-particle duality stands as the main example of the Bohr's complementarity principle, which states that quantons have properties that are equally real but mutually exclusive \cite{Bohr}. For instance, in a two-way interferometer, such as the Mach-Zehnder interferometer or the double-slit interferometer, the wave property is coded in the visibility of the interference pattern, while the particle nature is captured by the which-way information of the path along the interferometer, such that the complete manifestation of the one destroys the appearance of the other. The possibility for partial-simultaneous manifestation of the wave and particle aspects of a quanton was first quantitatively formulated and discussed by Wootters and Zurek \cite{Wootters}, and later it was expressed in terms of a simple complementarity relation between predictability $(P)$ and visibility $(V)$ \cite{Yasin}: 
\begin{align}
  V^2 + P^2 \le 1.  \label{eq:cr}
\end{align}
Such relation implies that these characteristics of a quanton are not necessarily mutually exclusive, i.e., it is possible to have a partial manifestation of the wave and particle nature of a quanton in the same experimental setup such that the more information one has about one aspect of the system, the less information the experiment can provide about the other, as was verified experimentally in Ref. \cite{Serra}.

Until now, many paths were taken to describe quantitatively the wave-particle aspect of a quanton \cite{ Engle, Ribeiro, Bera, Bagan, Coles, Hillery, Maziero, Leopoldo}. For instance, the extension of such complementarity relations for $d$-dimensional quantons was only possible with the realization that the quantum coherence \cite{Baumgratz} can be considered as the natural generalization for the visibility of a qubit \cite{Bera, Bagan, Mishra, Tabi}. As well, in Ref. \cite{Maziero}, it was shown that the mathematical properties of the density matrix of a quanton $A$ leads to several complementarity relations of the type
\begin{equation}
    C(\rho_{A})+P(\rho_{A})\le c(d_{A}), \label{eq:cr1}
\end{equation}
where $c(d_{A})$ is a constant that depends only on the system $A$ dimension $d_A$, $C(\rho_{A})$ is a quantum coherence measure and $P(\rho_{A})$ is a corresponding predictability measure, with both satisfying the criteria established in Refs. \cite{Durr, Englert} for bone-fide measures of visibility and predictability. Besides, the complementarity relations of the type derived in Refs. \cite{Yasin, Maziero} saturate if, and only if, the quantum state of the quanton $A$ is pure. 

For mixed states, the complementarity relation never saturates and the left hand side of Eq. (\ref{eq:cr1}) can even reach zero for a maximally mixed state. As noticed by Jakob and Bergou \cite{Janos}, due to the purification theorem, a maximally mixed state of the quanton $A$ can be seen as being due to its maximal entanglement with another quantum system $B$. Triality relations, also known as complete complementarity relations (CCRs), involving predictability, visibility and entanglement, were first suggested by Jakob and Bergou provided that a bipartite quantum system $AB$ is in a  pure state. Taking the purity of a bipartite  quantum system as the main hypothesis, two of us  derived CCRs of the type \cite{Marcos}:
\begin{align}
    C_{re}(\rho_A) + P_{vn}(\rho_A) + S_{vn}(\rho_A) = \log_2 d_A. \label{eq:ccr1}
\end{align}
 Here, $C_{re}(\rho_A) := S_{vn}(\rho_{A diag}) - S_{vn}(\rho_{A})$ is the relative entropy of quantum coherence, with $\rho_{A diag}$ being the diagonal part of $\rho_{A}$, $P_{vn}(\rho_A) = \log_{2}d_{A} - S_{vn}(\rho_{A diag})$ is the corresponding bone-fide predictability measure and $S_{vn}(\rho_A) = - \Tr \rho_A \log_2 \rho_A$ is the von Neumann entropy, which is an entanglement monotone. It is noteworthy that the predictability and entanglement measures can be seen as a measure of path distinguishability in an interferometer, as recently discussed in Refs. \cite{Qureshi, Wayhs}. Besides, Eq. (\ref{eq:ccr1}) has astonishing aspects that were shown recently: it is intrinsically connected to the notion of contextual realism defined by Bilobran and Angelo \cite{Bilobran}, as discussed in Ref. \cite{Jonas}; it is invariant by global unitary operations, which implies that it is preserved under unitary evolution and it is Lorentz invariant \cite{CCRin}; it remains valid in curved spacetimes as a quanton travels along its world lines, as shown in Ref. \cite{CCRcst}. 

In this work, our goal is twofold. We simulate a biased Mach-Zehnder interferometer $(MZI)$ in a circuit-based quantum computer, where both the beam splitters are not balanced and discuss such experiment in the light of complementarity relations, what leads to some novelties. For instance, in the biased $MZI$, we discuss situations where the interferometric visibility does not reflect the behavior of quantum coherence inside the interferometer.
It is noteworthy that, in Refs. \cite{Mishra, Tabi}, the authors also discussed situations where the interferometric visibility is not a good measure for the wave aspect for cases where the dimension of the quanton $A$ is bigger than two and argued that the quantum coherence is a better measure of the waviness of the quantum system with dimension $d_A > 2$. More specifically, in Ref. \cite{Mishra} the authors made a theoretical analysis of the experimental observations reported in Ref. \cite{Mei} in a four-path quantum interference experiment via multiple-beam Ramsey interferometry. In this experiment, a selective scattering of photons from just one interfering path causes decoherence and, at the first glance, it seems that there is an increase in the contrast of the interference pattern, i.e., of the interferometric visibility. However, the authors highlighted the fact that the visibility fails to capture the wave nature of the quanton in this multi-path experiment and argued that quantum coherence  remains a good quantifier of wave nature in such situations. Therefore, the authors concluded that the enhancement of the visibility in the presence of environmental decoherence stresses the limitations of the traditional measure of interferometric visibility as a quantifier for the wave character of the quanton. 

In contrast, in this article, we present some cases where the visibility is not a good measure as the quantum coherence for quantifying the wave property even for $d_A = 2$. And this is a consequence of the experimental setup we use, not relying in the behavior of the interferometric visibility in the presence of decoherence, i.e., the interaction of the quanton with an environment. Therefore, in our case, the fact that the visibility fails to catch completely the wave behavior is due to its own traditional standard definition in the context of interferometry. To the best of our knowledge, this is the first article showing that the usual definition of visibility fails to quantifying completely the wave aspect of a quanton for $d_A=2$. In the cases explored here, the visibility can be taken, at most, as an indicative of partial wave-behavior of the quanton inside the interferometer. As well, we discuss the fact that the complementarity relations based on the quantum coherence as a measure of the waviness are better since the quantifiers of the wave-particle aspect of a quantum system depend only on the state of the system inside the interferometer. On the other hand, the definition of interferometric visibility is made with regard to the detection probabilities, thus making reference to the state of system after the second beam splitter, which in this case is also a biased beam splitter. Therefore, the interferometric visibility uses the state of the quanton outside the interferometer to measure the wave aspect of the quanton inside the interferometer, and this, as we show here, leads to some unexpected results in the case of a biased $MZIs$.

The remainder of this article is organized as follows. In Sec. \ref{sec:gmzi}, we use a biased Mach-Zehnder interferometer (BMZI) to argue that interferometric visibility is not, in general, a good quantifier for the wave character of a quanton. In Sec. \ref{sec:ibmq}, we test our findings experimentally by using IBM quantum computers to do a full quantum simulation of the optical biased MZI. Finally, in Sec. \ref{sec:conc}, we give our concluding remarks.

\section{Coherence vs visibility in a biased Mach-Zehnder interferometer}
\label{sec:gmzi}

\begin{figure}[t]
    \centering
    \includegraphics[scale=0.3]{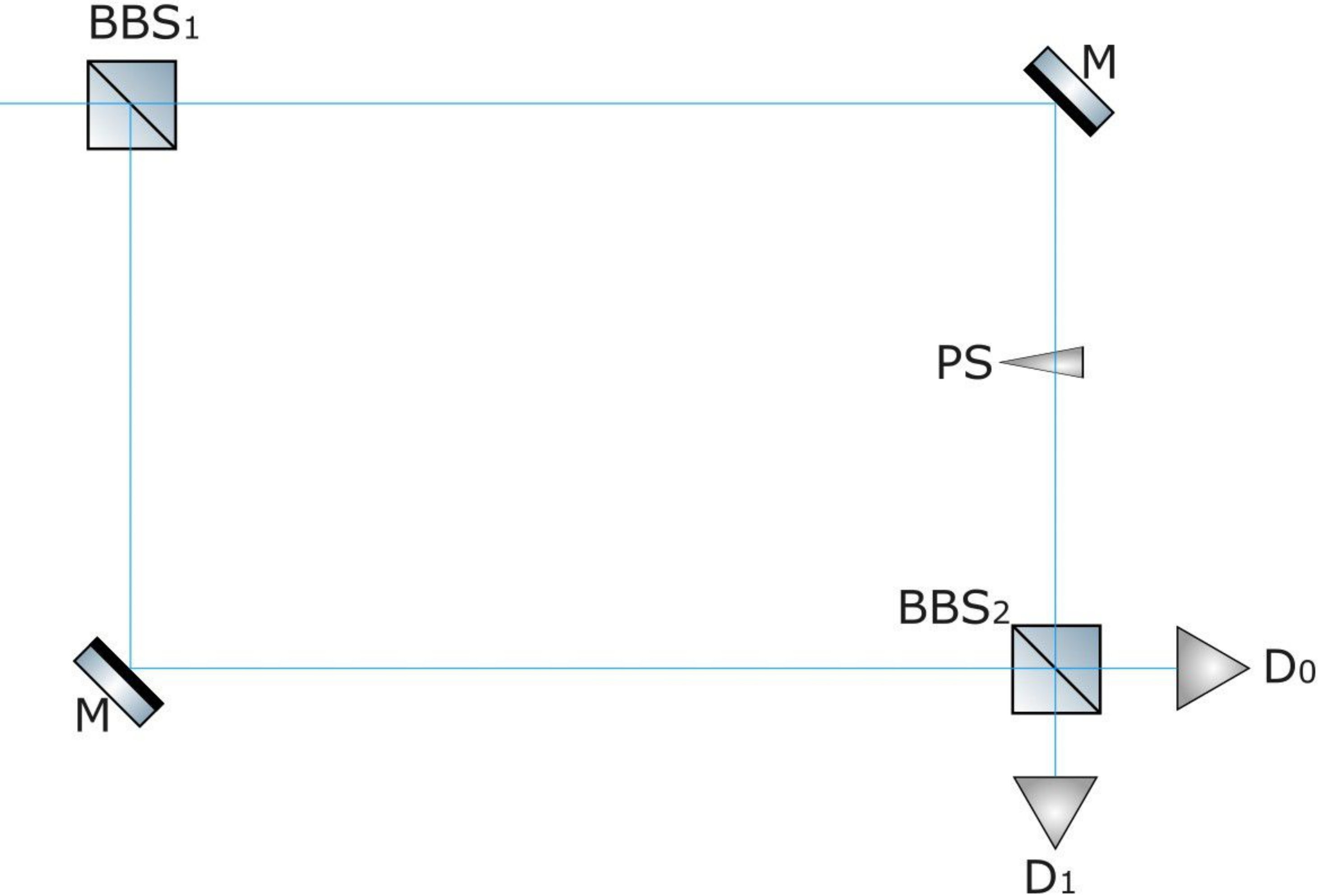}
    \caption{Schematic representation of the biased Mach-Zehnder interferometer. $BBS_{j}$ represents the biased beam splitters, $M$ stands for the mirrors, $PS$ is the phase-shifter, and $D_{j}$ are the photon detectors.}
    \label{fig:machzeh}
\end{figure}

In contrast with many articles and textbooks on Quantum Mechanics, that consider two balanced beam splitters $(BS)$ or one balanced BS \cite{Ellis, Auletta, Liu, Lu}, in this article we regard an experimental setup for the Mach-Zehnder interferometer with two biased beam splitters ($BBSs$) characterized by arbitrary transmission $\left(  T_j\right)  $ and reflection $\left(  R_j\right)$ coefficients, with both coefficients being real numbers between $0$ and $1$ such that $T_j^{2}$ is the probability of quantons being transmitted in the $j$-beam splitter, while $R_j^{2}$ is the probability of quantons being reflected in the beam splitter $j = 1,2$, as depicted in Fig. \ref{fig:machzeh}. Therefore, we must have $T_{j}^{2}+R_{j}^{2}=1.$ It is noteworthy that the authors in Refs. \cite{Oster, Calva} already considered such general setup for didactic purposes. 

Besides, from optics, we have that each reflection causes a phase shift of $\frac{\pi}{2}$, which is equivalent to a phase shift of $e^{i \frac{\pi}{2}} = i$ in the wave function. 
We will use the following notation for the states of the spatial mode of the quanton: $\ket{0} := \ket{h}$ represents the horizontal spatial mode whereas $\ket{1} := \ket{v}$ stands for the vertical spatial mode. The quanton $A$ is initially prepared in the state $\ket{\psi_0} = \ket{0}$. After the first $BBS$, the state of the quanton is given by $\ket{\psi_1} = T_1 \ket{0} + i R_1 \ket{1}.$
After the mirrors and phase-shifter, we have
\begin{align}
    \ket{\psi_2} = i e^{i \phi} T_1 \ket{1} - R_1 \ket{0}. \label{eq:psi2}
\end{align}
With the action of the second $BBS$, the state of system is given by
\begin{align}
    \ket{\psi_3}  =& -\big(e^{i \phi} T_1 R_2 + R_1 T_2 \big)\ket{0} + i\big(e^{i \phi}T_1 T_2 - R_1 R_2 \big) \ket{1}. \label{eq:psi3}
\end{align}
From Eq. (\ref{eq:psi3}), we use 
Born's rule to obtain the probabilities for the detectors $D_{0}$ and $D_1$ to click:
\begin{align}
\Pr\left(  D_{0}\right) & = T_{1}^{2}R_{2}^{2}+R_{1}^{2}T_{2}^{2}+2T_{1}R_{1}T_{2}R_{2}\cos\phi, \label{eq:pr0} \\
\Pr\left(  D_{1}\right) & = T_{1}^{2}T_{2}^{2}+R_{1}^{2}R_{2}^{2}-2T_{1}R_{1}T_{2}R_{2}\cos\phi. \label{eq:pr1}
\end{align}

By using the usual definition of interferometric visibility as the contrast of the interference pattern found in textbooks \cite{Auletta} or in the works of Greenberger and Yasin \cite{Yasin} and Englert \cite{Engle}, a closer examination of Eqs. (\ref{eq:pr0}) and (\ref{eq:pr1}) shows that we can have different expressions for the interferometric visibility depending on which detector we use to define such a measure:
\begin{align}
    \mathcal{V}_0 & := \frac{\Pr(D_0)_{\max}-\Pr(D_0)_{\min}}{\Pr(D_0)_{\max}+\Pr(D_0)_{\min}} = \frac{2T_{1}R_{1}T_{2}R_{2}}{T_{1}^{2}R_{2}^{2}+R_{1}^{2}T_{2}^{2}}, \label{eq:vis0} \\
    \mathcal{V}_1 & := \frac{\Pr(D_1)_{\max}-\Pr(D_1)_{\min}}{\Pr(D_1)_{\max}+\Pr(D_1)_{\min}}  = \frac{2T_{1}R_{1}T_{2}R_{2}}{T_{1}^{2}T_{2}^{2}+R_{1}^{2}R_{2}^{2}}.\label{eq:vis1}
\end{align}

It is straightforward to see that if the second $BBS$ is balanced, i.e., $T_2 = R_2 = 1/\sqrt{2}$ and $T_1 = T, R_1 = R$, as was the case considered in Ref. \cite{chen}, both measures of interferometric visibility coincide and we have the usual expression $\mathcal{V}_0 = \mathcal{V}_1 = \mathcal{V} = 2 TR$, which is equal to $C_{l_1}(\ketbra{\psi_2})$, the $l_1-$norm quantum coherence \cite{Baumgratz}. 

\begin{figure}[t]
    \centering
    \subfigure[$\mathcal{V}_0$ as a function of $T_1$ and $T_2$.]{{\includegraphics[scale = 0.48]{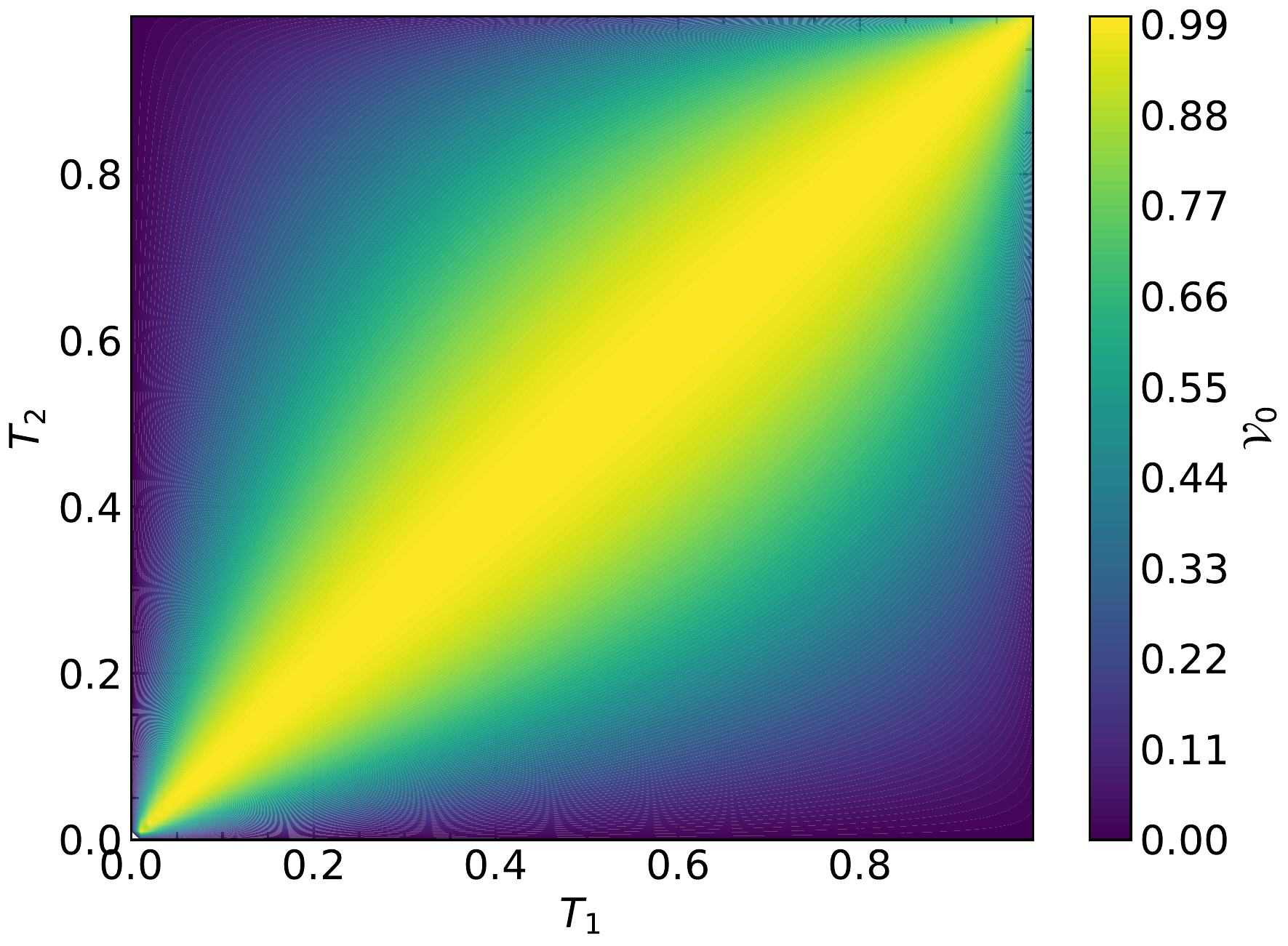}{\label{fig:vis}} }}
    \subfigure[$\mathcal{V}_1$ as a function of $T_1$ and $T_2$.]{{\includegraphics[scale = 0.48]{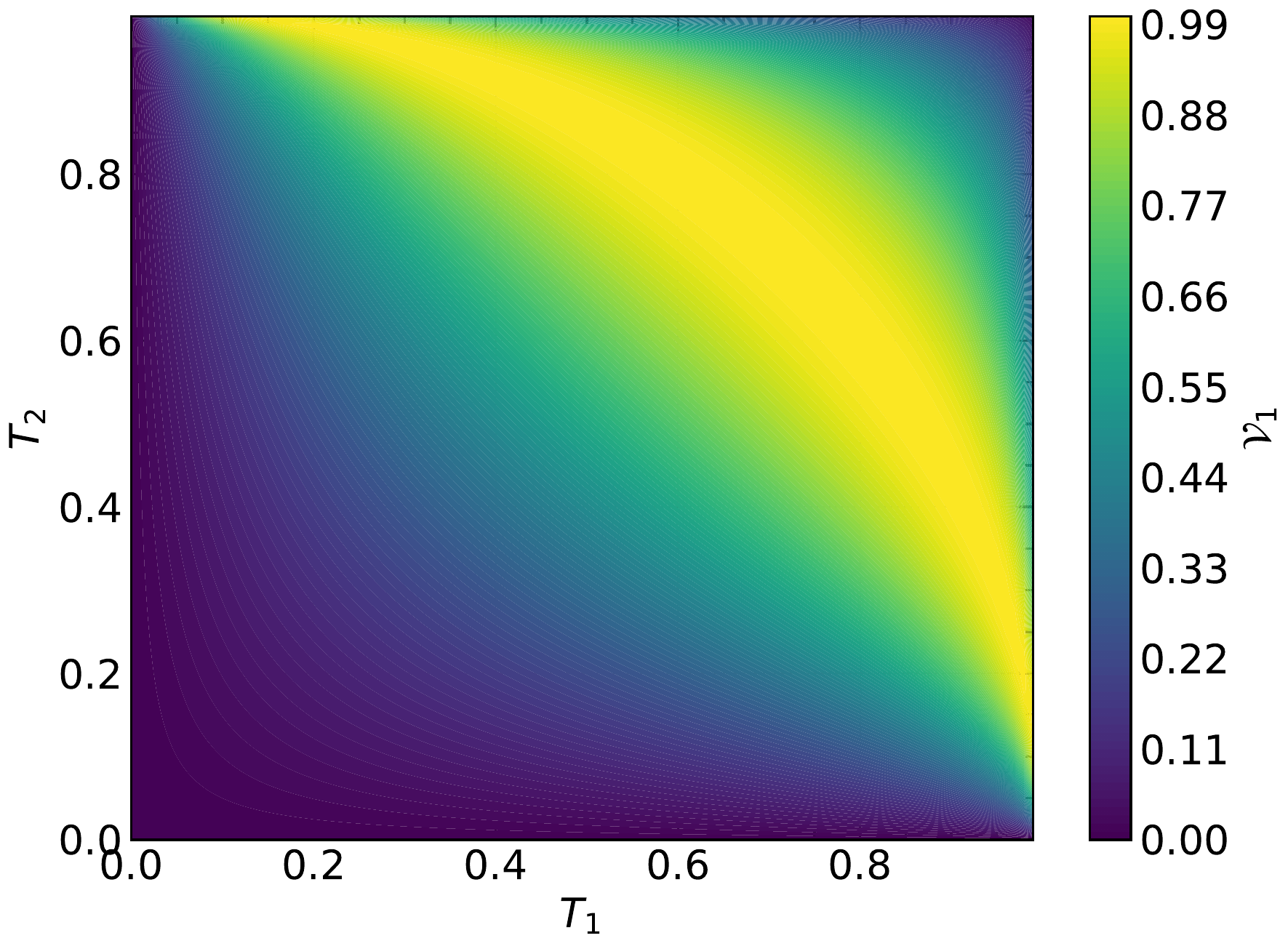}{\label{fig:vis1}} }}
  \caption{The interferometric visibilities $\mathcal{V}_0$ and $\mathcal{V}_1$ as a function of the transmission coefficients $T_1$ and $T_2$ of the biased beam splitters.} 
  \label{fig:visi}
\end{figure}

On the other hand, if $T_1 = R_1 = 1/\sqrt{2}$,  while $T_2 = T$ and $R_2 = R$ are free to vary, we have $\mathcal{V}_0 = \mathcal{V}_1 = \mathcal{V} = 2 TR$ but the visibility does not reflect the behavior of quantum coherence inside the interferometer, as we will see below. Moreover, in the most general case, where both beam-splitters are not balanced, we encounter two different expressions for the visibility depending on which detector we use. This is illustrated graphically in Fig. \ref{fig:visi}, where we plotted $\mathcal{V}_0$ and $\mathcal{V}_1$ as a function of $T_1$ and $T_2$. It is worth mentioning that $\mathcal{V}_0$ is not defined for the points $T_1 = T_2 = 0$ and $R_1 =  R_2 = 0$ whereas $\mathcal{V}_1$ is not defined for the points $T_1 = R_2 = 0$ and $R_1 =  T_2 = 0$ \cite{Oster}. However this is not a problem, since, for instance, if $T_1 = T_2 = 0$ the quanton always hit the detector $D_1$, and there's no interference pattern. Actually, we see that $\mathcal{V}_0 = \mathcal{V}_1$ if and only if $T_{1}=R_{1}$ or $T_{2}=R_{2}$. Besides that, the expressions for $\mathcal{V}_0$ and $\mathcal{V}_1$ depend on $T_2$ and $R_2$, i.e., on the transmission and reflection coefficients of the second $BBS$, instead of just on $T_1$ and $R_1$ as is expected for a quantity that should quantify the wave behavior of the quanton $A$ inside the interferometer, i.e., the state superposition with respect to the two arms of the interferometer. 

It is worth mentioning that here we are concerned with the definition of visibility in the quantum case, i.e., when we have only one quanton at the time inside the interferometer and a limited number of quantons in total. Besides, we are not dealing with the application of IVI in quantum metrology. In this article we are concerned only with its use for quantifying the wave character of a \textit{single} quantum system.

Now, without loss of generality, let us consider the expression for $\mathcal{V}_0$. A similar analysis can be done for $\mathcal{V}_1$. In Ref. \cite{Yasin}, the path predictability measure was introduced through a guessing game such that the smartest strategy would be to guess that all the particles will be in the same beam, namely the more intense one. The next step is then to compare the success of this guess with that obtained in the case where we have no success at all, i.e., if both paths are equally likely. If we do this here, we will end up with the same expression proposed by Greenberger and Yasin \cite{Yasin}, i.e., $P = |T_1^2 - R_1^2|$. Although $P$ is a bone-fide measure of predictability, $P$ and $\mathcal{V}_0$ do not form a complementarity relation for a pure state, as should occur, i.e., $P^2 + \mathcal{V}^2_0 \neq 1$. In this scenario, of the MZI with two $BBS$, a predictability measure that accompanies $\mathcal{V}_0$ is the following one
\begin{align}
    \mathcal{P}_0 = \frac{\abs{R_1^2 T_2^2 - R_2^2 T_1^2}}{R_1^2 T_2^2 + R_2^2 T_1^2}.
\end{align}
For this function we have
\begin{align}
    \mathcal{P}_0^2 + \mathcal{V}_0^2 = 1. \label{eq:cr2}
\end{align}
Here, we have an equality instead of an inequality once the state of system is pure. Besides, if the second $BBS$ is balanced, it is easy to see that $\mathcal{P}_0 = P$ and the complementarity relation given by Eq. (\ref{eq:cr2}) reduces to Eq. (\ref{eq:cr}), obtained in Ref. \cite{Yasin} for the pure state case. Again, we see that $\mathcal{P}_0$ is also a function of $T_2$ and $R_2$, what is not reasonable since $\mathcal{P}_0$ should reflect just the particle aspect of the quanton $A$ inside the interferometer.

In order to reinforce that these measures for predictability and visibility do not reflect what is expected in the \textit{quantum realm}, let us consider the following case: \textbf{(a)}
Let us suppose that $T_1 = T_2 = T$ and $R_1 = R_2 = R$. In this case, we have that $\mathcal{V}_0 = 1$ and $\mathcal{P}_0 = 0$, independent of the values of $T$ and $R$, which can be considered as a pathological case. For instance, if $T \gg R$ or $R\gg T$, we should expect that $\mathcal{P}_0 \gg \mathcal{V}_0$ and not the other way around. And, certainly, we should not expect  $\mathcal{V}_0$ to reach its maximum possible value while $\mathcal{P}_0 \approx 0$ when there is almost no path superposition (quantum coherence) inside the interferometer. The explanation for $\mathcal{V}_0$ to be independent of the values of $T$ and $R$ comes from the fact that\footnote{In the cases where $\Pr(D_j)_{\min}=0$, we have $\mathcal{V}_j=(\Pr(D_j)_{\max}-\Pr(D_j)_{\min})/(\Pr(D_j)_{\max}+\Pr(D_j)_{\min})=1$, for $j = 0$ or $1$, even for $P(D_j)_{\max}\ll 1$. So, if we have a limited number of quantons to experiment with, we see that $\mathcal{V}_j'=\mathcal{V}_j\Pr(D_{j})_{\max}$ would be a better quantifier for the visibility of the interference fringes contrast. It is worthwhile though noticing that this change would not solve the problems IVI has regarding the quantification of the wave character of a quanton.} $\Pr(D_0)_{\min} = 0$, which implies that $\mathcal{V}_0 = \Pr(D_0)_{\max}/\Pr(D_0)_{\max} = 1$. Of course, in the classical case, the interference pattern and the visibility is sharp since we are dealing with the intensity of light instead of probabilities. However, in the quantum realm, if we have a limited number of photons and one photon at the time is inside the interferometer, as nowadays the technology already permits, and if $T \gg R$, then $\Pr(D_0)_{\max} \ll 1$, we know beforehand that the predictability of the path is high and the quantum coherence of \textit{single} quanton is small inside the interferometer. Therefore, here the inteferometric visibility can be taken, at most, as a indicative of partial wave behavior of \textit{single} quanton inside the interferometer. Besides, one could argue that $\mathcal{V}_1$ should be used instead of $\mathcal{V}_0$ in this situation, but this is an arbitrary choice. Similarly, for $T_1 = R_2$ and $R_1 = T_2$, then $\mathcal{V}_1 = 1$ and the same analysis follows for $\mathcal{V}_1$. Therefore, here we show a situation where the definition of the visibility is not a good measure for the wave aspect of the quanton inside the interferometer. In this situation, $\mathcal{V}_0$ and $\mathcal{V}_1$ will correctly reflect the wave behavior of the quanton only if both $BBS$ are balanced, i.e., $T_1 = R_1 = T_2 = R_2$.

A more interesting case, that will be addressed below in a more general setting, where $\mathcal{V}_0$ is not a bone-fine measure of the waviness of the quanton, is when $T_1 = R_1$ such that $\mathcal{V}_0 = \mathcal{V}_1 = 2 T_2 R_2$, i.e., even though both interferometric visibilities coincide, $\mathcal{V}_0$ and $\mathcal{V}_1$ depend on $T_2$ and $R_2$ while we have full superposition of the quanton inside the interferometer. A similar behavior appears in experimental settings usually used in quantum delayed choice experiments (QDCE) where a balanced second $BS$ is prepared in a superposition of being in and out of the interferometer, as in Ref. \cite{Terno, Ma}. However, in Ref. \cite{Dieguez} the authors discussed the fact that the visibility at the output has no connection whatsoever with the wave element of reality, as defined in accordance with the criterion of realism of a given observable (in this case, a `wave'-observable), as defined in Ref. \cite{Bilobran}. Besides, they proposed a setup that removes this objection and establishes a monotonic link between the visibility and wave elements of reality inside the interferometer. To do this, the authors considered the second $BS$ balanced and the first $BS$ in a superposition of being in and out of the interferometer. We just mentioned this work here because we want to point out the fact that, in the case where the quanton is initially prepared in the state $\ket{\psi_0} = \ket{0}$, the IVI  $\mathcal{V}_0$ will be a bone-fide measure of the wave aspect of system only when the second $BBS$ is balanced. In this case, both interferometric visibilities coincide and $\mathcal{V}_0 = \mathcal{V}_1  = 2 T_1 R_1 = C_{l_1}(\ketbra{\psi_2})$.

It is noteworthy that, in Refs. \cite{Mishra, Tabi}, the authors also discussed situations where the visibility is not a good measure for the wave aspect for cases where the dimension of the quanton is bigger than two. They argued that the quantum coherence is a better measure of the wave aspect of the quanton system with dimension $d_A > 2$. In contrast, in this article we report a situation where the visibility is not a good measure for the wave aspect of a quantum system even for dimension $d_A = 2$, where the usual definition of visibility should work. This is a consequence of the experimental setup we used, which makes the definition of the visibility inappropriate for a \textit{single} quanton inside the interferometer since $\Pr(D_0)_{min} = 0$ or depends on $T_2$ and $R_2$.

On the other hand, if we consider the CCR expressed by Eq. (\ref{eq:ccr1}) applied to this scenario, since the quanton is prepared in a pure state, then $S_{vn}(\rho_A) = 0$ and Eq. (\ref{eq:ccr1}) reduces to
\begin{align}
     C_{re}(\rho_A) + P_{vn}(\rho_A) = 1. \label{eq:cr3}
\end{align}
Once such relation is invariant under unitary transformations, it is preserved under unitary evolution and it is applicable in every step of the biased $MZI$. For instance, inside the interferometer, after the mirrors and the phase shifter, $C_{re}(\rho_A), P_{vn}(\rho_A)$ will be functions only of $T_1$ and $R_1$, as expected since $\rho_A$ will be the density operator regarding the state of the system inside the interferometer, i.e., regarding the state given by Eq. (\ref{eq:psi2}). In contrast, the definition of the visibility is regarding the detection probabilities and thus making reference to the state of system after the second $BBS$, i.e., regarding the state given by Eq. (\ref{eq:psi3}). Therefore, the visibility uses the state of the quanton outside the interferometer to measure the wave aspect of the quanton inside the interferometer. This, as we showed above, leads to some problems in the case of the biased $MZI$. In contrast, the complementarity relation given by Eq. (\ref{eq:cr3}) uses the state of the quanton inside the interferometer to quantify the wave and particle aspects of the system:
\begin{align}
    & C_{re}(\ketbra{\psi_2}) = - R_1^2 \log_2 R_1^2 - T_1^2 \log_2 T_1^2,\label{eq:cretr} \\
    & P_{vn}(\ketbra{\psi_2}) = 1 + R_1^2 \log_2 R_1^2 + T_1^2 \log_2 T_1^2,\label{eq:pvntr}
\end{align}
with both measures being a function only of $T_1$ and $R_1$, as expected. 

\begin{figure}[t]
    \centering
    \includegraphics[scale=0.25]{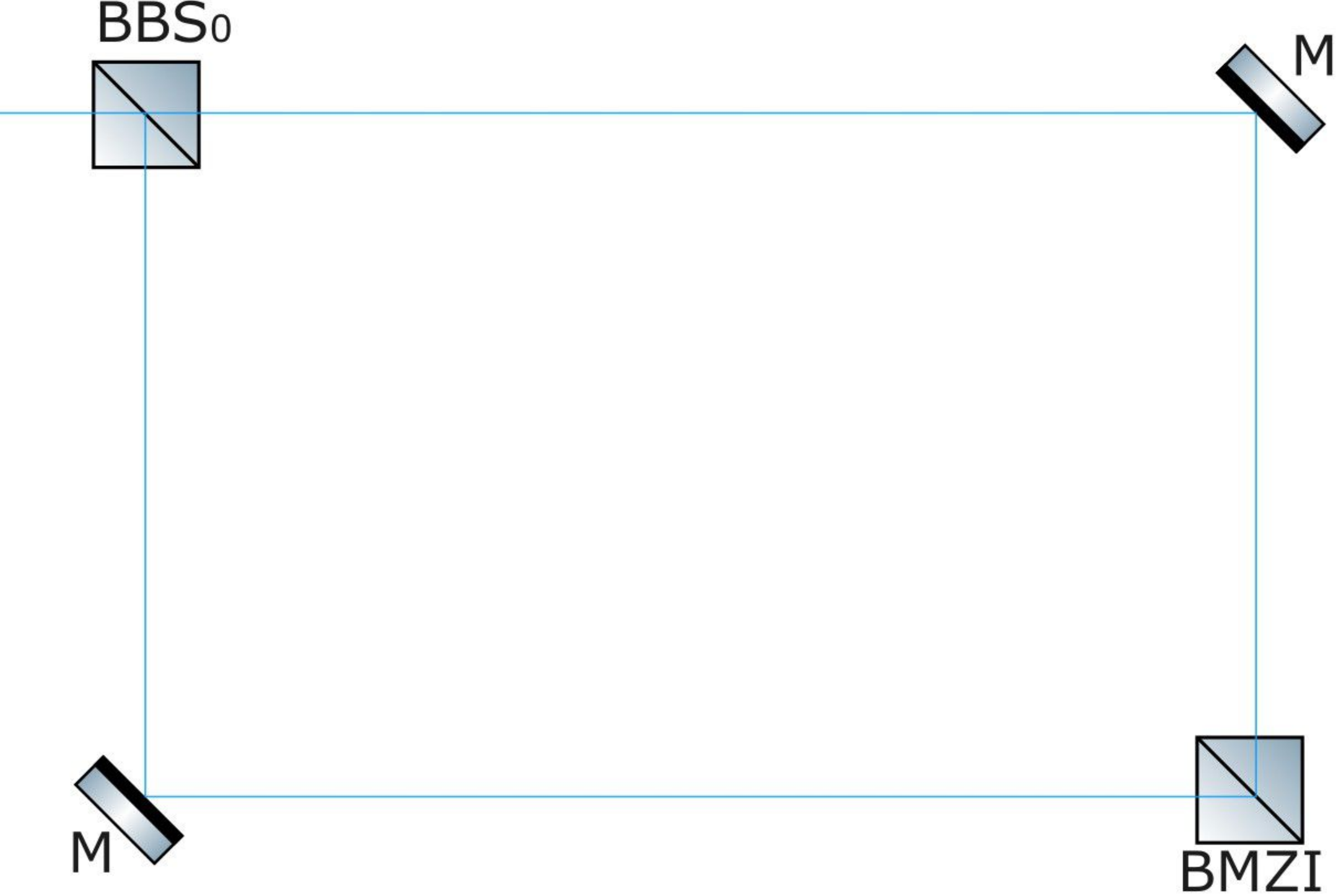}
    \caption{Schematic representation of the double biased Mach-Zehnder interferometer. $BMZI$ is the biased Mach-Zehnder interferometer illustrated in Fig. \ref{fig:machzeh}.}
    \label{fig:doubmachzeh}
\end{figure}

Now, let us consider the following case:
\textbf{(b)} 
A quanton is prepared in a general pure state $\ket{\psi_0} = \alpha \ket{0} + \beta \ket{1}$, with $\abs{\alpha}^2 + \abs{\beta}^2 = 1$, that goes through the biased $MZI$. In order to prepare such state, we just have to consider another biased-$MZI$ before $BBS_1$, as depicted in Fig. \ref{fig:doubmachzeh}. In this case, following the same steps as before, the probability of detection in $D_{0}$ is given by:%
\begin{align}
\Pr\left(  D_{0}\right) =& \abs{\alpha}^2\left(  T_{1}^{2}R_{2}%
^{2}+R_{1}^{2}T_{2}^{2}\right) +\abs{\beta}^2\left(  T_{1}^{2}T_{2}%
^{2}+R_{1}^{2}R_{2}^{2}\right) \nonumber \\
&  +2\left(\abs{\alpha}^2-\abs{\beta}^2\right)  T_{1}R_{1}%
T_{2}R_{2}\cos\phi  -2\left(  \operatorname{Re}\alpha\operatorname{Re}\beta+\operatorname{Im}%
\alpha\operatorname{Im}\beta\right)  R_{2}T_{2}\sin\phi  \\
&  +2\left(  \operatorname{Im}\alpha\operatorname{Re}\beta-\operatorname{Re}%
\alpha\operatorname{Im}\beta\right)\left[  R_{1}T_{1}\left(  R_{2}^{2}%
-T_{2}^{2}\right)  +\left(  R_{1}^{2}-T_{1}^{2}\right)  R_{2}T_{2}\cos
\phi\right], \nonumber
\end{align}%

A similar expression follows for $\Pr(D_1)$, but we choose not to show it here because it does not add anything new to the present analysis. One can see that the visibility will be a function of $R_1, T_1, R_2, T_2$, in general. Besides, with $\alpha=1,\ \beta=0$ one can easily see that the case explored before is recovered. Now, it is interesting to notice that when we choose $\alpha=\beta =1/\sqrt{2}$, the probability for $D_{0}$ reduces to
$\Pr\left(D_{0}\right)  = \frac{1}{2} - R_2 T_2 \sin \phi,
$
which implies that the visibility is given by $\mathcal{V}_0 = \mathcal{V}_1 = 2T_{2}R_{2}$, i.e., the visibilities coincide, however they are only a function of $R_2$ and $T_2$. For instance, if the first $BBS$ is balanced, i.e., $T_1 = R_1 = 1/\sqrt{2}$, the state of system right after the $BBS_1$ is $\ket{\psi_1}= \frac{1+i}{2}(\ket{0} + \ket{1})$. So, after the mirrors and phase-shifter the state is given by $\ket{\psi_2} = \frac{i - 1}{2}(\ket{0} + e^{i \phi}\ket{1})$, which implies that we have maximal superposition (or maximal quantum coherence) and therefore the wave-behavior inside the interferometer. However, $\mathcal{V}_j$ depends on $T_2$ and $R_2$ and will quantify the waviness of the system only if $T_2 = R_2 = 1/\sqrt{2}$ once in this case $\mathcal{V}_j = C_{l_1}(\ketbra{\psi_2}) =1$. In contrast, if $BBS_2$ is not balanced, the interferometric visibility will not quantify correctly the wave aspect of the quanton inside the interferometer. It is worth observing that, in this case, the fact the $\mathcal{V}_j$ is not a bone-fide measure comes directly from the fact that $\mathcal{V}_j$ depends on $T_2$ and $R_2$ and not from $\Pr(D_0)_{\min} = 0$. Here we have the opposite behavior between quantum coherence and interferometric visibility in comparison with case \textbf{(a)}. In the case \textbf{(a)} the quantum coherence varies while the visibility $\mathcal{V}_0$ is constant, whereas in \textbf{(b)}, when the first $BBS$ is balanced, the quantum coherence is constant and the visibility varies. On the other hand, if we use the complementarity relation given by Eq. (\ref{eq:cr3}), we have that $P_{vn}(\ketbra{\psi_2}) = 0$ and $C_{re}(\ketbra{\psi_2}) = 1$ independent of $R_2$ and $T_2$, as expected.


\section{Full quantum simulation of the biased Mach-Zehnder interferometer using IBM's quantum computers}
\label{sec:ibmq}
 
\begin{figure}[t]
    \centering
    \includegraphics[scale=0.9]{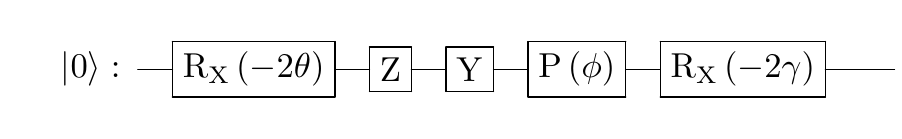}
    \caption{The quantum circuit implemented in IBMQ for simulating the Mach-Zehnder interferometer with biased beam splitters ($BBS$'s). The first  $R_X(-2\theta)$ gate (left to the right) is the $BBS_1$. For the mirrors, it is necessary to apply the $Z$ gate and then the $Y$ gate. $P(\phi)$ is the phase-shifter and, finally, the second $R_X(-2\gamma)$ represents the $BBS_2$.}
    \label{fig:gmzi_ibmq}
\end{figure}

In this section, we simulate the biased $MZI$ using IBM's quantum computers \cite{IBMQ}, in order to provide a proof-of-principle experimental verification of our theoretical results. To do this, we simulate the optical elements of the biased $MZI$ through unitary gates as follows. The $BBS$ is implemented by
$
U_{BBS}(\theta)  = R_X(-2\theta) = \left[
\begin{array}
[c]{cc}%
\cos\theta & i\sin\theta\\
i\sin\theta & \cos\theta
\end{array}
\right],
$
where $R_X\left(  \theta\right)$ is a rotation by an angle $\theta$ about the $X$ axis. 
So, we use $T:=\cos\theta$ and $R:=\sin\theta$, with
$\theta\in\left[0,\pi/2\right]$. As a particular case, we can get the balanced $BS$ by
setting $\theta=\pi/4$. The mirrors' combined action is implemented using
$
YZ = \left[
\begin{array}
[c]{cc}%
0 & i\\
i & 0
\end{array}
\right],
$
where $Y$ e $Z$ are the usual Pauli matrices. Therefore, one can see that mirrors give a phase shift of $e^{i\frac{\pi}{2}}$ in the state vector and the vertical and horizontal path states are switched. The phase shift is implemented through the phase gate $P(\phi) = \ketbra{0} + e^{i \phi} \ketbra{1}$. Finally, we need to apply the $U_{BS}(\gamma)=RX(-2\gamma)$.
\begin{figure}[t]
    \centering
    \includegraphics[scale=0.7]{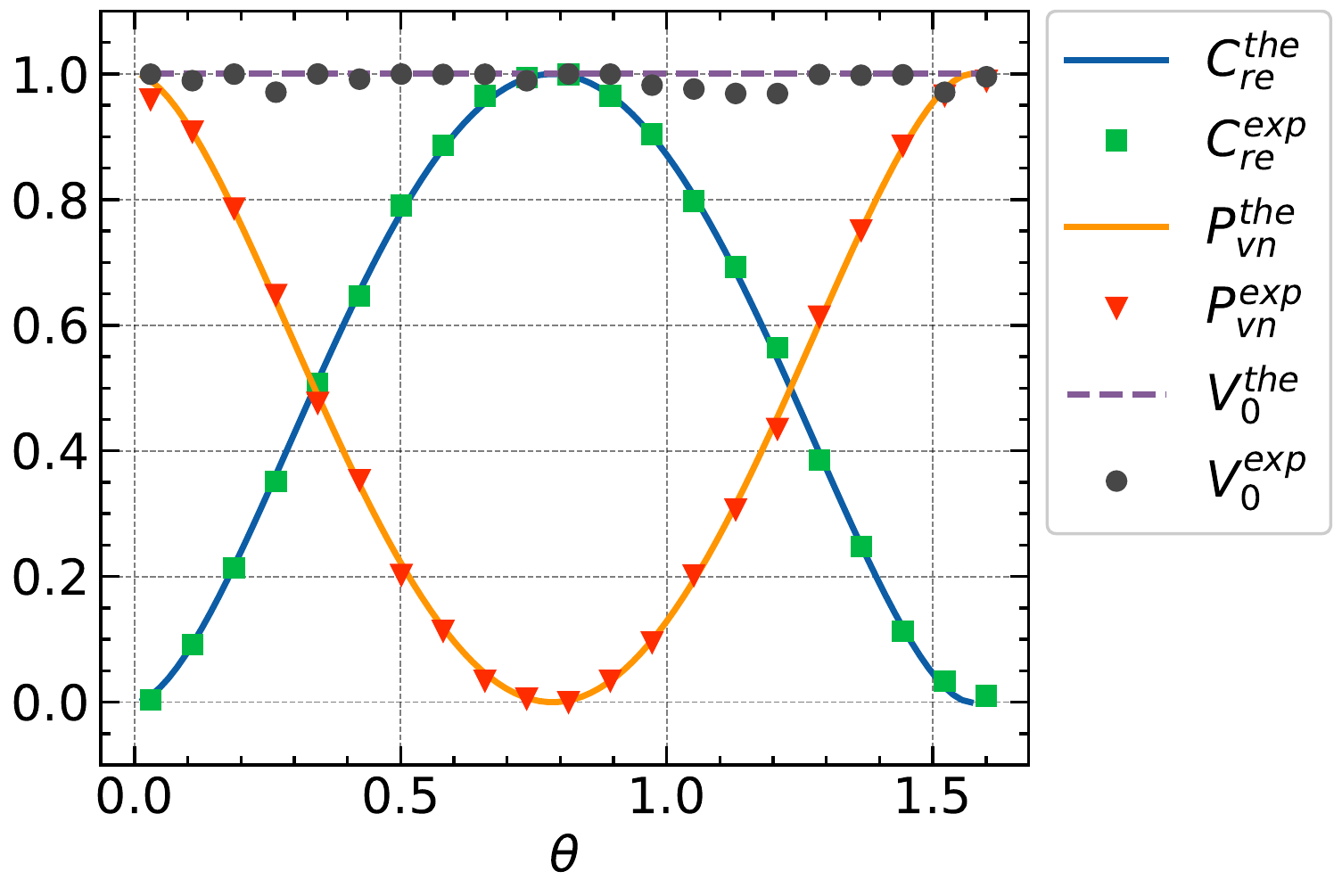}
    \caption{Theoretical (lines) and experimental (points) results for quantum coherence and predictability of $|\psi_{2}\rangle$ and visibility of $|\psi_{3}\rangle$ for the case with the input state being $\ket{0}$, $T_1 = T_2 = T = \cos \theta$ and $R_1 = R_2 = R = \sin \theta$ with $\theta \in[0,\pi/2]$.}
    \label{fig:exp_t1_t2_e}
\end{figure}

\begin{table}
\caption{\label{armonk} Calibration parameters for the Armonk chip version 2.4.33.}
\begin{tabular}{l c}
\hline 
 Armonk parameters & Q0  \tabularnewline
\hline 
\hline 
Frequency (GHz) & 4.797  \tabularnewline
T1 ($\mu$s) & 272.47  \tabularnewline
T2 ($\mu$s) & 276.02  \tabularnewline
Gate error ($10^{-1}$) & 1.56 \tabularnewline
Readout error ($10^{-2}$) & 4.06  \tabularnewline
\hline
\end{tabular}
\label{table}
\end{table}

To exemplify our results, we implement the circuit for the biased $MZI$ shown in Fig. \ref{fig:gmzi_ibmq} and use the Armonk quantum chip of IBM's Quantum Experience, whose calibration parameters are shown in Table \ref{table}. Between the phase-shifter and the $BBS_2$ we performed a quantum state tomography to obtain information of the density matrix $\rho_{2}=|\psi_{2}\rangle\langle\psi_{2}|$, from which we calculate the quantum coherence and predictability using Eqs. (\ref{eq:cretr}) and (\ref{eq:pvntr}), respectively. A second state tomography is performed after the application of $BBS_2$, so that we can calculate the visibilities of Eqs. (\ref{eq:vis0}) and (\ref{eq:vis1}) through the maximum and minimum probabilities by the variation of the phase $\phi$.
For the input state $\ket{0}$ and for $T_1 = T_2 = T = \cos \theta$ and $R_1 = R_2 = R = \sin \theta$ with $\theta \in[0,\pi/2]$, the results are shown in Fig. \ref{fig:exp_t1_t2_e}.
It is important to mention that we used the Qiskit tools for measurement error mitigation \cite{qiskit}, which improved substantially the experimental results.
As one can see, for $T\gg R$ and $R\ll T$,  the quantum coherence of the state $\rho = \ketbra{\psi_2}$ inside the interferometer has small values and the predictability is high, while $\mathcal{V}_0 = 1$ independent of the values of $T$ and $R$.

\section{Conclusions}
\label{sec:conc}

For a fairly long time, interferometric visibility (IVI) has been considered as a synonymous of the wave character of a quantum system (WCQ). Recently, indications were given, using decoherence, that quantum coherence is more suited than IVI for quantifying the WCQ of a \textit{single} quanton. In this article, we used a biased Mach-Zehnder interferometer (BMZI) to settle this issue, even for a two-dimensional quantum system. We identified sets of parameters of the BMZI for which IVI indicates quite the opposite of what one knows to be the behavior of a \textit{single }quanton inside the interferometer. This is in a way expected, since IVI is computed using the quanton's state outside of the interferometer and we want to quantify the WCQ inside the interferometer.

\begin{acknowledgments}
This work was supported by the Universidade Federal do ABC (UFABC), process 23006.000123/2018-23, by the 
Coordena\c{c}\~ao de Aperfei\c{c}oamento de Pessoal de N\'ivel Superior (CAPES), process 88887.649600/2021-00, by the Conselho Nacional de Desenvolvimento Cient\'ifico e Tecnol\'ogico (CNPq), process 309862/2021-3, and by the Instituto Nacional de Ci\^encia e Tecnologia de Informa\c{c}\~ao Qu\^antica (INCT-IQ), process 465469/2014-0.
\end{acknowledgments}

\vspace{0.3cm}

\textbf{Data availability} The Qiskit code used for implementing the simulations and experiments used to obtain the data utilized in this article is available upon request to the authors.

\end{document}